\documentclass[sigconf]{acmart}

\usepackage{booktabs} % For formal tables

\copyrightyear{2017} 
\acmYear{2017} 
\setcopyright{othergov}
\acmConference{ThematicWorkshops'17}{October 23--27, 2017}{Mountain View, CA, USA}\acmPrice{15.00}\acmDOI{10.1145/3126686.3126735}
\acmISBN{978-1-4503-5416-5/17/10}

\begin{document}

\title{Learning Transferable Features for Speech Emotion Recognition}
%\titlenote{Produces the permission block, and
%  copyright information}
%\subtitle{Extended Abstract}
%\subtitlenote{The full version of the author's guide is available as
%  \texttt{acmart.pdf} document}

\author{Alison Marczewski}
%\authornote{First author}
\orcid{0000-0001-9249-5351}
\affiliation{%
  \institution{CS Dept., UFMG \& Kunumi}
 \city{Belo Horizonte}
 \state{Brazil}
}
\email{amarczew@dcc.ufmg.br}

\author{Adriano Veloso}
%\authornote{The secretary disavows any knowledge of this author's actions.}
\orcid{0000-0002-9177-4954}
\affiliation{%
  \institution{CS Dept., UFMG}
  \city{Belo Horizonte}
  \state{Brazil}
}
\email{adrianov@dcc.ufmg.br}

\author{Nivio Ziviani}
%\authornote{First author}
\orcid{0000-0002-7079-2010}
\affiliation{%
	\institution{CS Dept., UFMG \& Kunumi}
	\city{Belo Horizonte}
	\state{Brazil}
}
\email{nivio@dcc.ufmg.br}

\renewcommand{\shortauthors}{A. Marczewski et al.}

\begin{abstract}
Emotion recognition from speech is one of the key steps towards emotional intelligence in advanced human-machine interaction. Identifying emotions in human speech requires learning features that are robust and discriminative across diverse domains that differ in terms of language, spontaneity of speech, recording conditions, and types of emotions. This corresponds to a learning scenario in which the joint distributions of features and labels may change substantially across domains. In this paper, we propose a deep architecture that jointly exploits a convolutional network for extracting domain-shared features and a long short-term memory network for classifying emotions using domain-specific features. We use transferable features to enable model adaptation from multiple source domains, given the sparseness of speech emotion data and the fact that target domains are short of labeled data. A comprehensive cross-corpora experiment with diverse speech emotion domains reveals that transferable features provide gains ranging from 4.3\% to 18.4\% in speech emotion recognition. We evaluate several domain adaptation approaches, and we perform an ablation study to understand which source domains add the most to the overall recognition effectiveness for a given target domain.
\end{abstract}




\maketitle

\section{Introduction}

Humans are increasingly interacting with machines via speech, which is an important impetus for studying the vocal channel of emotional expression. Applications of an interface capable of assessing emotional states from human voice are numerous and diverse, including communication systems for vocally-impaired individuals, call centers, lie detection, airport security, and realistic interaction with empathy. The aim of this work is the development of models capable of recognizing people's emotions from recorded voice, also known as emotion recognition from speech.

Most emotional states involve physiological reactions, which in turn modify different aspects of the voice production process~\cite{juslin}. Emotions produce changes in respiration and an increase in muscle tension, which influence the vibration of the vocal folds and vocal tract shape, thus affecting the acoustic characteristics of the speech. When someone is in a state of anger, fear or joy, the sympathetic nervous system is aroused, the heart rate and blood pressure increase, the mouth becomes dry and there are occasional muscle tremors. As a result, speech is loud, fast and enunciated with strong high frequency energy. Sadness, by contrast, is associated with a low, hesitant, and lacking in energy speech~\cite{feat2}.

While there is considerable evidence that speech features can differentiate emotional states~\cite{affect4,affect3,affect2}, the way in which physiological reactions translate into speech features may vary greatly depending on specific factors such as acoustic signal conditions, speakers, spoken languages, linguistic content, and type of emotion (e.g., acted, elicited, or naturalistic)~\cite{spont}. Since each possible combination of such factors may define a specific domain, emotion recognition from speech becomes particularly challenging because it is unclear which speech features are the most effective for each domain. Also, it is challenging to train an emotion recognition system exclusively for the target domain due to unavailability of sufficient labeled data which limits the exploration of the feature space. Fortunately, there are potentially shared or local invariant features that shape emotions in different domains, thus transfer learning may alleviate the data demands.

In this paper, we propose a deep architecture for speech emotion recognition composed of a convolutional neural network (CNN)
%to extract domain-shared features from multi-domain data,
and a long short-term memory network (LSTM).
%that is fed with the extracted domain-shared features for emotion prediction and uses a limited amount of target-domain data.
The main hypothesis in this work is that the blend of a CNN with a LSTM exploits both spatial and temporal information of speech features for emotion recognition. That is, while the CNN extracts spatial features of varying abstract levels, the LSTM employs contextual information in order to model how emotions evolve over time. We discuss several feature transference approaches designed to our deep architecture. Such feature transference approaches differ in terms of the choice of which layers to freeze or tune, and whether or not target domain data are used during pre-training.

We conducted rigorous experiments using six standard speech emotion datasets that correspond to different domains. Recognition models are trained using different transference approaches, and we pose the following questions:
\begin{itemize}
\item Which feature transference approach is the most appropriate, given factors such as the amount of labels and the discrepancy between domains?%Which are prone to co-adaptation, source specialization, and overfitting?
\item How effective is the blend of CNN with LSTM networks for domain adaptation?
\item How effective is our recognition model compared with the state-of-the-art models for speech emotion recognition based on supervised domain adaptation?
\end{itemize}

We performed an ablation domain analysis in order to elucidate the benefits of incorporating multi-domain data into the final recognition model. We show that even small amounts of multi-domain data used for adaptation can significantly improve recognition effectiveness, while domain discrepancy poses serious issues to effective model adaptation. Also, the effectiveness of the different feature transference approaches varies greatly depending on the factors that define the target domain. We report gains that vary from 4.3\% to 18.4\%, depending on the target domain and feature transference approach.

%leads to negative transference, that is, when transferred features lead to a lower performance on the target task than the no-transference approach.

%However, an important simplification is that the factors that define each domain are known apriori. Therefore, we may take advantage of processing each audio sample in a particular way depending on the corresponding domain. Thus, we pre-train a CNN with dynamic freezing, a novel learning algorithm which controls the specificity of the extracted features based on the discrepancy between source and target domains. If source and target domains are discrepant, then most of the CNN layers are frozen and their weights are not updated. On the other hand, high-level or specific features are also extracted.

%Figure~\ref{fig:domain} depicts this intuition. Domain-specific features tend to produce large separations and low error within the domain, but they also tend to produce poor models with regard to discrepant domains. By contrast, while generic features do not prioritize separation, they tend to be meaningful to other domains.

%\begin{figure*}[t]
%\centering
%\includegraphics[scale=0.45]{tt2.eps}
%\includegraphics[scale=0.45]{tt1.eps}
%\includegraphics[scale=0.45]{tt3.eps}
%\caption{(Color online) Left $-$ Source domains $d_1$ and $d_2$ employ domain-specific features in order to increase separability and decrease error. Right $-$ Domain $d_1$ employs generic features because it is discrepant from the target domain.}
%\label{fig:domain}
%\end{figure*}

\section{Related Work}

Research on the recognition of emotional expressions in voices is of great academic interest in psychology~\cite{psych2}, neurosciences~\cite{neuro1,neuro2,neuro3,neuro4} and affective computing~\cite{affect7,affect6,affect4,affect3}. A number of researchers investigated acoustic correlates of emotions from human speech. In one of the first studies~\cite{feat3}, the authors identify parameters in the speech that reflect the emotional state of a speaker. They found that anger, fear, and sorrow situations tend to produce characteristic differences in contour of fundamental frequency, average speech spectrum, temporal characteristics, precision of articulation, and waveform regularity of successive glottal pulses.

\vspace{0.05in}
\noindent{\bf{Features}} There are studies on how acoustic correlates of emotions from speech are transformed into features for supervised learning algorithms. In~\cite{review,review2}, the authors provide reviews on a wide range of features employed for emotion recognition from speech. In~\cite{eurospeech}, the authors present an approach based on hidden semi-continuous Markov models, which are built using specific energy and pitch features. In~\cite{gaussian}, the authors employ mel frequency cepstral coefficients (MFCCs) as features for a Gaussian mixture model classifier. A similar MFCC model was proposed in~\cite{rate} and features related to speaking rate are also explored to categorize the emotions. In~\cite{prosodic}, the authors propose speech prosody and related acoustic features for the recognition of emotion. Methods for emotion recognition from speech relying on long-term global prosodic features were developed. In~\cite{emotion1}, the authors describe seven  acoustic and four linguistic types  of features, from which they found the most important ones, and also discuss the mutual influence of acoustics and linguistics. In~\cite{emotion4}, the authors introduce string kernels as a novel solution in the field.

\vspace{0.05in}
\noindent{\bf{Data Concerns}} Background noise, varying recording levels, and acoustic properties of the environment, and how these issues impact speech emotion recognition systems are discussed in~\cite{affect1}. More serious concerns about data used for emotion recognition from speech were presented in~\cite{affect6}, where the authors discuss issues related to the overestimation of the accuracy of emotion recognition systems, since experiments are usually performed on acted data (rather than on spontaneous data). Concerns with experiments performed on acted data were also discussed in~\cite{emotion3}. Alternatively, more realistic acted data were recently presented in~\cite{corpus}.

\vspace{0.05in}
\noindent{\bf{Transfer Learning and Domain Adaptation}} Since speech data are usually captured from different scenarios, it is often observed a significant performance degradation due to the inherent mismatch between training and test set. Thus, domain adaptation is a relevant topic in emotion recognition from speech. In~\cite{cross2}, the authors explore a multi-task framework in which speech or song are jointly leveraged in emotion recognition in a cross-corpus setting. In~\cite{cross3}, the authors show that training and test data used for system development usually tend to be similar as far as recording conditions, noise overlay, language, and types of emotions are concerned. The authors conclude that a cross-corpus evaluation would provide a more realistic view of the recognition performance. In~\cite{domain1}, the authors propose a feature transfer approach using a deep architecture called PCANet, which extracts both the domain-shared and the domain-specific latent features, leading to significant effectiveness improvements. In~\cite{domain2}, the authors propose a two-layer network, so that the parameters within the second layer are imposed the common priors between the related classes, so that the classes with few labeled data in target domain can borrow knowledge from the related classes in source domain. In~\cite{domain4}, the authors present a feature transfer learning method using denoising autoencoders~\cite{autoencoder} to build high order sub-spaces of the source and target corpora, where features in the source domain are transferred to the target domain by a specific neural network. Similarly, in~\cite{affect4}, the authors employ a denoising autoencoder as a domain adaptation method. In this case, prior knowledge learned from a target set is used to regularize the training on a source set. Finally, in~\cite{supdomain}, the authors propose a supervised domain adaptation approach which can improve the speech emotion recognition performance in the presence of mismatched training and testing conditions. In~\cite{baseline} the authors propose feature transfer learning based on sparse autoencoders. Their approach consists of learning a representation using a single-layer autoencoder, and then applying a linear SVM using the learned representation.

\vspace{0.05in}
\noindent{\bf{Feature Learning}} Deep neural networks were already used for emotion recognition from speech. In~\cite{affect2}, the authors propose a generalized discriminant analysis using deep neural networks. They show that low-dimensional features capture hidden information from the acoustic features leading to  significant gains compared with typical SVMs. In~\cite{affect8}, the authors assume a scenario where speech data are obtained from different devices and varied recording conditions. As a result, data are typically highly dissimilar in terms of acoustic signal conditions. They evaluate the use of denoising autoencoders~\cite{autoencoder} to minimize this data mismatch problem. In~\cite{deep1}, the authors propose the use of deep neural networks to extract high level features from raw recorded voice. The network outperforms SVMs using hand-crafted features. In~\cite{deep2}, the authors employ deep belief networks and their results suggest that learning high-order non-linear relationships using these networks is an effective approach for emotion recognition. In~\cite{autoencoder3}, the authors employ a feature enhancement method based on an autoencoder with LSTMs, for robust emotion recognition from speech. The enhanced features are then used by SVMs. In~\cite{cnn3}, the authors propose to learn salient features for speech emotion recognition using CNNs. The network is learned in two stages. In the first stage, unlabeled samples are used to learn local invariant features using sparse autoencoders with reconstruction penalization. In the second step, these features are used as the input to a feature extractor. In~\cite{dd}, the authors introduce an approach to separate emotion-specific features from general and less discriminative ones. They employ an unsupervised feature learning framework to extract rough features. Then these rough features are further fed into a semi-supervised feature learning framework. In this phase, efforts are made to disentangle the emotion-specific features and some other features by using a novel loss function, which combines reconstruction penalty, orthogonal penalty, discriminative penalty and verification penalty.

\vspace{0.05in}
\noindent{\bf{Our Work}} The main differences between this work and aforementioned works are: (i) we consider diverse domain adaptation approaches using CNN and LSTM features, (ii) we perform a domain ablation analysis which reveals the relative value of different domains, (iii) we perform domain blending, that is, we not just transfer features from one domain to another, but we produce generic features using data from multiple domains simultaneously. Further, we investigated the best freezing/tuning cut-off for each target domain.

\section{Multi-Domain Network}

The task of learning to recognize emotions from speech is defined as follows. We have as input the {\em training set} (referred to as $\mathcal{D}$), which consists of a set of records of the form $<a,e>$, where $a$ is an audio sample (i.e., an emotional episode) and $e$ is the corresponding emotion being expressed. Emotions draw their values from a discrete set of possibilities, such as sadness, fear, happiness, surprise, and anger. The training set is used to construct a model which relates features within the audio samples to the corresponding emotions. The {\em test set} (referred to as $\mathcal{T}$) consists of records $<a,?>$ for which only the audio sample $a$ is available, while the corresponding emotion $e$ is unknown. The model learned from the training set $\mathcal{D}$ is used to produce estimations of the emotions expressed on audio samples in the test set $\mathcal{T}$.

We consider a learning scenario in which audio samples and their corresponding emotion labels are drawn from different generating distributions. For instance, some audio samples may be obtained from acted speech while other audio samples are obtained from spontaneous speech. The process that produces audio samples may also differ in terms of factors such as recording conditions, spoken language, and linguistic content. A specific combination of these factors defines a {\em domain}. Speech emotion recognition is a domain-specific problem, that is, a recognition model learned from one domain is likely to fail when tested against data from another domain~\cite{domain5}. As a result, real application systems usually require labeled data from multiple domains, guaranteeing an acceptable performance for different domains. However, each domain has a very limited amount of labels due to the high cost %associated with creating
to create large-scale labeled datasets for domain-specific speech emotion recognition. Feature transferability is thus an appealing way to alleviate the demands for domain-specific labels. Thus, for domains that are short of labeled data transferable features enable model adaptation from multiple domains. 

\subsection{Network Architecture}

\begin{figure}[ht]
\centering
\includegraphics[scale=0.3]{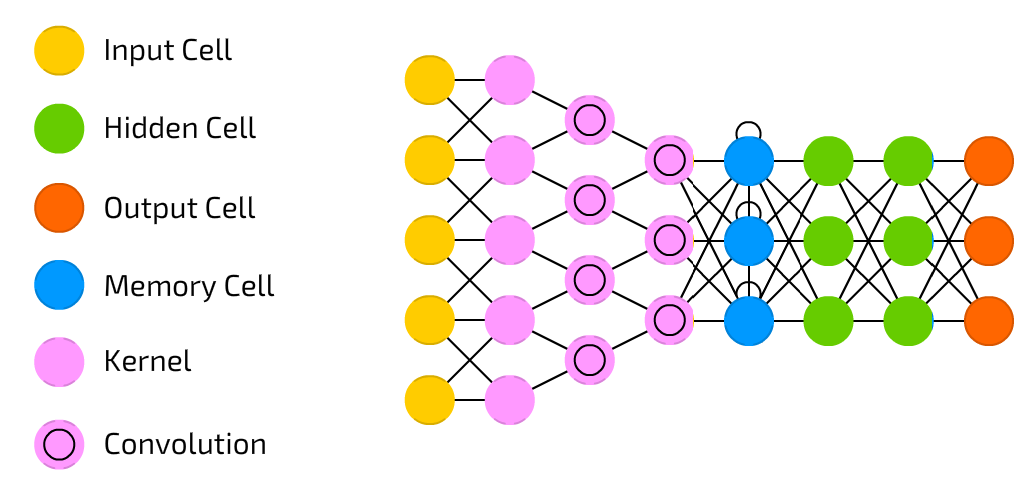}
\caption{(Color online) Multi-Domain Network architecture for learning transferable features. Convolutional layers are followed by a LSTM layer. Different feature transference approaches are designed using this architecture.}
\label{fig:net1}
\end{figure}

In this section, we introduce our deep architecture. It first extracts generic features from multi-domain data (or domain-shared features) which are then used to produce domain-specific and highly discriminative features. The architecture combines a deep hierarchical spatial feature extractor with a model that can learn to recognize and synthesize temporal dynamics of emotions, as illustrated in Figure~\ref{fig:net1}. The network works by passing each audio sample through a feature transformation to produce a fixed-length vector representation.\footnote{Padding was applied so that audios with different durations have representations with the same size. Also, features are standardized so that they are centered around 0 with a standard deviation of 1.} After that, spatial features are computed for the audio input, and then the sequence model captures how emotions evolve over time.

More specifically, the network receives a 54,000 dimensional input representing audio samples. It has five hidden layers, including two uni-dimensional convolutional layers, one LSTM layer, and two fully connected layers. The convolutional layers apply kernels with 128 dimensions, combined with ReLUs and a dropout level of 0.30. The LSTM layer receives 128 dimensional inputs, and returns two 500 dimensional vector outputs which are then flattened into a single 1,000 dimensional output. The next two fully connected layers are composed of 1,000 units and are combined with the hyperbolic tangent activation. Again, a dropout level of 0.30 is applied. The final classification layer employs a softmax cross-entropy loss and thus the minimization problem is given as: 

\begin{equation}
\mbox{min} \frac{1}{n}\displaystyle\sum_{i=1}^{n} J(\theta(x_i), y_i)
\nonumber
\label{eq:ce}
\end{equation}

\noindent where $J$ is the cross-entropy loss function and $\theta(x_i)$ is the conditional probability that the network assigns $x_i$ to emotion label $y_i$. 
The network is trained by the AdaDelta method, and six emotions are considered, namely: anger, disgust, fear, happiness, sadness, and surprise. The network architecture is substantially smaller than others commonly used. We also evaluated deeper networks, but the resulting models showed to be less accurate and learning becomes significantly slower.

\begin{table*} [ht!]
\caption{Summary of the datasets.}
\centering
  \begin{tabular}{lccccccc} \hline
	  Dataset/Domain & Age    & Language & Emotion & Gender    & Recording & Sampling rate & \# samples\\\hline\\
	  AFEW       & children/adults & English & natural & balanced & movies & 48kHz & 568\\
	  Emo-DB     & adults & German   & acted   & balanced   & studio    & 16kHz & 287\\
	  EMOVO      & adults & Italian & acted & balanced & studio & 48kHz & 336\\
	  eNTERFACE  & adults & English  & induced & unbalanced  & normal    & 16kHz & 1,047\\
	  IEMOCAP    & adults & English & acted & balanced & studio & 48kHz & 1,770\\
	  RML        & adults & many & induced & balanced & studio & 22kHz & 650\\
  \hline
  \end{tabular}
\label{tab:data}
\end{table*}

\subsection{Feature Transferability}

We assume the presence of few labeled audio samples in the target domain, hence a direct adaption to the target domain via fine-tuning is prone to overfitting. We also assume that the training set is composed of audio samples belonging to different domains, and we can explicitly split $\mathcal{D}$ into $n$ different domains, that is, $\mathcal{D}=d_1, d_2, \ldots, d_n$. Thus, the goal of our deep architecture is to train a multi-domain network to differentiate emotions based on input audios associated with multiple domains. Although audio samples associated with a given domain $d_i$ may be better represented by specific features, there still exist some common features that permeate all other domains. Examples of such low-level features may include pitch, derivative of pitch, energy, derivative of energy, duration of speech segments, among others.

%Intuitively, the network uses lower layers for modeling a domain which is discrepant from the target domain. Through this learning procedure, domain-independent information is modeled in the shared layers from which useful generic feature representations are obtained.

\vspace{0.05in}
\noindent{\bf{Transference Approaches}} The main intuition that we exploit for feature transferability is that the features must eventually transition from general to specific along our deep architecture, and feature transferability drops significantly in higher layers with increasing domain discrepancy~\cite{bengio2}. In other words, the features computed in higher layers must depend greatly on a specific domain $d_i$, and recognition effectiveness suffers if $d_i$ is discrepant from the target domain. Since we are dealing with many domains simultaneously, we also considered multiple transference approaches, which are detailed next:
\begin{itemize}
\item A1: no fine-tuning is performed, which means that the pre-trained model is used to recognize emotions.
\item A2: no layer is kept frozen during fine-tuning, which means that errors are back-propagated through the entire network during fine-tuning.
%\footnote{Freezing a layer means not changing the weights of the layer during Stochastic Gradient Descent optimization.}
\item A3: only the first convolutional layer is kept frozen during fine-tuning.
\item A4: both convolutional layers are kept frozen during fine-tuning.
\item A5: convolutional and LSTM layers are kept frozen during fine-tuning. That is, errors are back-propagated only through the fully-connected layers during fine-tuning.
\item A6: only the first convolutional layer is kept frozen during fine-tuning. All other layers have their weights randomly initialized for fine-tuning.
\item A7: both convolutional layers are kept frozen during fine-tuning. All other layers have their weights randomly initialized for fine-tuning.
\item A8: convolutional and LSTM layers are kept frozen during fine-tuning. Weights in fully-connected layers are randomly initialized for fine-tuning.
\end{itemize}

Further, these transference approaches are applied considering different scenarios:
\begin{itemize}
\item S1: target domain data are used during pre-training and fine-tuning.
\item S2: target domain data are used exclusively during fine-tuning.
\end{itemize}

\section{Experimental Results}

In this section, we present the datasets and baselines used to evaluate our multi-domain network for speech emotion recognition. Then we discuss our evaluation procedure and report the results of our multi-domain network. 

In particular, our experiments aim to answer the following research questions:
\begin{enumerate}
\item [RQ1] How effective is the blend of CNN with LSTM networks for speech emotion recognition? How do the learned features compare against hand-crafted features?
\item [RQ2] Which feature transference approach is more appropriate to each target domain?
\item [RQ3] Which domain characteristics affect the most the accuracy of the model?
\item [RQ4] How effective is our multi-domain compared with other domain adaptation models?
\end{enumerate}

\subsection{Datasets and Domains}

Our analysis is carried on six datasets which differ mainly in terms of language, number of speakers, number of emotions and spontaneity of speech. 
The details about each dataset are given next:
\begin{itemize}
\item AFEW~\cite{afew}: The Acted Facial Expressions In The Wild dataset contains segments from 37 movies in English. The movies have been chosen keeping in mind the need for different realistic scenarios and large age range of subjects to be captured.
\item Emo-DB~\cite{emodb}: The Berlin Emotional Speech dataset features actors speaking emotionally defined sentences. The dataset contains emotional sentences from 10 different actors and ten different texts.
\item EMOVO~\cite{emovo}: The dataset consists of sentences recorded by six professional actors. Each speaker reads fourteen Italian sentences expressing different emotions.
\item eNTERFACE~\cite{enterface}: The dataset consists of recordings of naive subjects from fourteen nations speaking pre-defined spoken content in English. The subjects listened to six successive short stories eliciting a particular emotion.
\item IEMOCAP~\cite{iemocap}: The Interactive Emotional Dyadic Motion Capture dataset features ten actors performing improvisations in English, specifically selected to elicit emotional expressions. Each sentence is labeled by at least three human annotators.
\item RML:\footnote{\url{http://www.rml.ryerson.ca/rml-emotion-database.html}} The  dataset contains audiovisual emotional expression samples that were collected at Ryerson Multimedia Lab. The RML emotion database is language and cultural background independent. The audio samples were collected from eight human subjects, speaking six different languages (English, Mandarin, Urdu, Punjabi, Persian, Italian). Different accents of English and Chinese were also included.
\end{itemize}

\begin{table} [htb!]
\caption{UAR numbers for different models. No domain adaptation is performed.}
\centering
  \begin{tabular}{lccc} \hline
Dataset   & SVM$-$IS & CNN & CNN$+$LSTM \\\hline\\
AFEW      & .333 & .344 & .338 \\
Emo-DB    & .645 & .622 & .659 \\
EMOVO     & .411 & .440 & .459 \\
eNTERFACE & .456 & .419 & .454 \\
IEMOCAP   & .719 & .673 & .684 \\
RML       & .482 & .581 & .631 \\
  \hline
  \end{tabular}
\label{tab:svm}
\end{table}

Table~\ref{tab:data} presents a summary of the datasets.
All datasets were normalized to cover the same emotional states. Specifically, we focus on the well-known six  emotions~\cite{cowie}: anger, disgust, fear, happiness, sadness, and surprise. 
\subsection{Baselines}

We considered the following methods in order to provide baseline comparison:

\begin{itemize}
\item SVM with Interspeech 2010 features (SVM$-$IS): the 1,582 acoustic features proposed in~\cite{inter} are fed into an SVM with RBF kernel~\cite{schuller2010interspeech}. The hyper-parameters of the SVM are chosen by cross-validation. The main objective of using this baseline is to answer RQ1.
\item Training on Target (TT): a model CNN$+$LSTM is trained using only the target domain data. No source domain data are used. The main objective of using this baseline is to assess the benefits of the different feature transference approaches.
\item Adaptive SVM~\cite{supdomain}: this is a supervised domain adaptation algorithm for speech emotion recognition. The approach poses an optimization problem which seeks a decision boundary close to that of an SVM trained from the source domain, while managing to separate the labeled data from the target domain.
\end{itemize}

\subsection{Setup}

We implemented our architecture using Keras~\cite{keras}. The measure used to evaluate the recognition effectiveness of our models is the standard Unweighted Average Recall (UAR),\footnote{The UAR metric is the sum of the recalls per class divided by the number of classes.} as presented in~\cite{inter}. We conducted five-fold cross validation where datasets are arranged into five folds with approximately the same number of audio samples each. At each run, four folds are used as training set and the remaining fold is used as test set. The results reported are the average of the five runs, and are used to assess the overall discrimination performance of the models. To ensure the relevance of the results, we assess the statistical significance of our measurements by means of a pairwise t-test~\cite{t-test} with p$-$value $\le 0.05$. 

\subsection{Results and Discussion}

The first experiment is concerned with RQ1. We present a comparison between SVM$-$IS trained with Interspeech 2010 features, and our deep architecture was trained with raw audio. We considered deep architectures with and without the LSTM layer to assess the impact of using both spatial and sequential features. Table~\ref{tab:svm} shows UAR numbers for the different models. For this experiment, no domain adaptation is performed. Instead, samples from all datasets were used for training and testing the models using five-fold cross-validation. On average, the CNN$+$LSTM model provides UAR numbers that are statistically superior than the numbers provided by SVM$-$IS and CNN models (which are statistically equivalent on average), except for the dataset AFEW. Thus, the features learned by CNN$+$LSTM architecture lead to significantly raised UAR numbers.

The next set of experiments is devoted to answer RQ2. We evaluate diverse feature transference approaches. Table~\ref{tab:adaptation} shows UAR numbers when our architecture is trained using solely target domain data (TT). Therefore, if the target domain is short on labeled data, the model will probably suffer from overfitting. The table also shows the gains obtained by each feature transference approach relatively to TT. That is, we investigated the best freezing/tuning cut-off for each target domain. On average, the best performing transference approach is S1$-$A2, which uses target domain data during pre-training and fine-tuning and no layer is kept frozen during fine-tuning. Further, gains tend to decrease as more layers are kept frozen during fine-tuning. However, the best approach varies greatly depending on the target domain.

Considering AFEW as the target domain, the best transference approaches are S1$-$A1, S1$-$A4, and S1$-$A7. Usually, using target domain data during pre-training is very beneficial, except for EMOVO for which the best performer was S2$-$A3.  Fine-tuning is extremely important in all cases, specially if target domain data are not used during pre-training. Gains for IEMOCAP are significantly lower than the gains obtained for other domains. Notice that IEMOCAP is the largest dataset, and thus TT achieves very high UAR numbers, which are hard to surpass with domain adaptation. For RML, the best transference approaches are those that freeze less layers. This is because RML is composed of highly diverse languages. Thus, freezing layers will only work if target domain data are used during pre-training. Otherwise, freezing layers would be clearly detrimental to domain adaptation. It is also important to mention that for each target domain, many feature transference approaches lead to significant improvements.

\begin{table*} [htb!]
\caption{Different feature transference approaches and scenarios. Numbers in bold indicate the highest gains for each target domain.}
\centering
\begin{tabular}{lc|rrrrrrrr | rrrrrrrr} \hline
	& UAR & \multicolumn{16}{c}{Gains over TT} \\
	&& \multicolumn{8}{c}{S1} & \multicolumn{8}{c}{S2}\\\cline{3-18}
Target    & TT   & A1    & A2    & A3    & A4    & A5    & A6    &  A7   & A8    & A1    & A2    & A3    & A4    & A5    & A6    & A7    & A8 \\\hline\\
AFEW      & .288 & \bf{.121}  & .101  & .047  & \bf{.120}  & .115  & .045  & \bf{.121}  & .116  & .042  & .015  & .024  & .052  & .019  & .007  & .103  & .029 \\
Emo-DB    & .614 & .047  & \bf{.117}  & .088  & .051  & .052  & .102  & .057 &  .086 & -.365  & .093  & .064  & .083  & .050  & .068  & .065  & .083  \\
EMOVO     & .518 & -.095 & .053  & .014  & -.061 & -.089 & -.071 &  .034 & .014  & -.372 & .093 & \bf{.109} & -.060 & -.041 & -.044  & .008 & -.017\\
eNTERF    & .441 & .032  & .133  & .114  & .061  & .032  & \bf{.153} &  .087 & .045  & -.353 & .027 & -.015  & -.016  & -.034  & .002  & -.026 & -.037 \\
IEMOCAP   & .682 & .004  & .004  & -.002 & -.009 & -.003 & .003 &  \bf{.017} & \bf{.017} & -.363 & .003 & -.015  & -.016  & -.034  & .003  & -.026 & -.035 \\
RML       & .623 & -.014 & .032  & .041  & .005  & -.002 & \bf{.073} & .035  & .028  & -.518 & \bf{.074} & .062 & -.085  & -.145  & .054  & -.087 & -.143 \\\hline
Average   & $-$ & .016 & .073 & .050 & .028 & .017 & .051 & .058 & .053 & -.321 & .051 & .038 & -.007 & -.031 & .015 & .006 & -.020\\
Std. & $-$ & .072 & .051 & .044 & .063 & .068 & .078 & .039 & .041 & .189 & .041 & .049 & .064 & .067 & .040 & .069 & .076\\
 
  \hline
  \end{tabular}
\label{tab:adaptation}
\end{table*}

The next set of experiments is devoted to answer RQ3. Table~\ref{tab:ablation} shows UAR numbers obtained with a domain ablation analysis. More specifically, the table shows UAR numbers obtained by different feature transference approaches after excluding one of the source domains from the pre-training. This enables us to grasp the domain characteristics that affect the most the effectiveness of our multi-domain network.

The reference UAR value (All) is given by the model built using data from all domains. We first analyze scenario S1, in which target domain data are used during pre-training and fine-tuning. As can be seen, in almost all cases it is better removing one of the source domains from pre-training. Using AFEW data during pre-training is highly detrimental in all cases. The probable explanation is that the AFEW domain is highly discrepant from all other domains. Similarly, IEMOCAP data are highly detrimental for AFEW, Emo-DB, eNTERFACE and RML target domains. IEMOCAP data are also very discrepant from other domains. Removing out-of-domain data from pre-training is not beneficial only for S1$-$A1 when RML is the target domain. Thus, we conclude that if target domain data are used during pre-training, it is detrimental to have out-of-domain data during pre-training, specially if out-of-domain data are highly discrepant from the target domain data.

Very different trends are observed when we analyze scenario S2. In this case, target domain data are used exclusively during fine-tuning, and therefore we may expect that out-of-domain data used during pre-training are less discrepant. Using IEMOCAP data during pre-training is highly beneficial. This is probable due to the size of IEMOCAP dataset. This is also a probable explanation for the robustness when removing specific out-of-domain datasets when IEMOCAP is the target domain. The RML domain seems to benefit the most from out-of-domain data. In general, we conclude that if target domain data are not included during pre-training, it is beneficial to have out-of-domain data during pre-training, even if out-of-domain data are highly discrepant from the target domain data.

\begin{table*} [htb!]
\caption{Domain ablation analysis. The table shows UAR numbers after excluding a domain from the pre-training, so a low UAR number indicates that an important domain was removed from pre-training. Symbol $\uparrow$ indicates that UAR has raised significantly. Symbol $\bullet$ indicates that UAR has not changed significantly. Symbol $\downarrow$ indicates that UAR has dropped significantly. We omitted UAR numbers for A5 to A8 in order to avoid cutter. Highest UAR numbers for each feature transference approach are highlighted in bold.}
\centering
  \begin{tabular}{ll|rrrr | rrrr} \hline
	  &               & \multicolumn{8}{c}{UAR numbers}\\
	  &               & \multicolumn{4}{c}{S1} & \multicolumn{4}{c}{S2}\\\cline{3-10}
	  Target  & Source        & A1   & A2  & A3 & A4 & A1 & A2 & A3 & A4 \\\hline\\
AFEW          & All        & .323 & .317 & .301 & .322 & .300 & .292 & \bf{.295} & .303 \\
	      & $-$ Emo-DB & .356$^\uparrow$ & .440$^\uparrow$ & .469$^\uparrow$ & .517$^\uparrow$ &  .304$^\uparrow$ & \bf{.306}$^\uparrow$ & \bf{.299}$^\uparrow$ & .283$^\downarrow$ \\
	      & $-$ EMOVO  & .380$^\uparrow$ & .442$^\uparrow$ & .473$^\uparrow$ & .561$^\uparrow$ &  .307$^\uparrow$ & .289$^\bullet$ & .262$^\downarrow$ & .322$^\uparrow$ \\
	      & $-$ eNTERFACE & \bf{.390}$^\uparrow$ & .464$^\uparrow$ & .487$^\uparrow$ & .566$^\uparrow$ &  .284$^\downarrow$ & .291$^\bullet$ & .269$^\downarrow$ & .314$^\uparrow$ \\
	      & $-$ IEMOCAP  & .315$^\bullet$ & \bf{.514}$^\uparrow$ & \bf{.572}$^\uparrow$ & \bf{.625}$^\uparrow$ & .237$^\downarrow$ & .272$^\downarrow$ & .280$^\downarrow$ & .275$^\downarrow$ \\
	      & $-$ RML    & .366$^\uparrow$ & .424$^\uparrow$ & .487$^\uparrow$ & .539$^\uparrow$ & \bf{.314}$^\uparrow$ & \bf{.298}$^\uparrow$ & .287$^\downarrow$ & \bf{.332}$^\uparrow$ \\
\hline\\
Emo-DB        & All        & .643 & .685 & .668 & .645 & .389 & \bf{.671} & .653 & .665 \\
	      & $-$ AFEW   & .725$^\uparrow$ & .830$^\uparrow$ & \bf{.835}$^\uparrow$ & \bf{.843}$^\uparrow$ & \bf{.397}$^\uparrow$ & .652$^\downarrow$ & .644$^\downarrow$ & .648$^\downarrow$ \\
	      & $-$ EMOVO  & .688$^\uparrow$ & .792$^\uparrow$ & .789$^\uparrow$ & .786$^\uparrow$ &.382$^\downarrow$ & \bf{.667}$^\bullet$ & .658$^\bullet$ & .638$^\downarrow$ \\
	      & $-$ eNTERFACE & .689$^\uparrow$ & .775$^\uparrow$ & .780$^\uparrow$ & .767$^\uparrow$ &.393$^\bullet$ & .620$^\downarrow$ & \bf{.668}$^\uparrow$ & .662$^\bullet$ \\
	      & $-$ IEMOCAP  & \bf{.798}$^\uparrow$ & \bf{.856}$^\uparrow$ & \bf{.840}$^\uparrow$ & .824$^\uparrow$ &.372$^\downarrow$ & .653$^\downarrow$ & .639$^\downarrow$ & .660$^\downarrow$ \\
	      & $-$ RML    & .692$^\uparrow$ & .762$^\uparrow$ & .748$^\uparrow$ & .778$^\uparrow$ &\bf{.401}$^\uparrow$ & .655$^\downarrow$ & .658$^\uparrow$ & \bf{.683}$^\uparrow$ \\
\hline\\
EMOVO         & All        & .469 & .545 & .525 & .496 &.325 & .566 & \bf{.574} & .487 \\
	      & $-$ AFEW   & \bf{.635}$^\uparrow$ & \bf{.691}$^\uparrow$ & \bf{.716}$^\uparrow$ & \bf{.735}$^\uparrow$ &.332$^\uparrow$ & .542$^\downarrow$ & \bf{.571}$^\bullet$ & \bf{.547}$^\uparrow$ \\
	      & $-$ Emo-DB & .566$^\uparrow$ & .664$^\uparrow$ & .675$^\uparrow$ & .664$^\uparrow$ &.320$^\bullet$ & .567$^\bullet$ & .549$^\downarrow$ & .528$^\uparrow$ \\
	      & $-$ eNTERFACE & .566$^\uparrow$ & .655$^\uparrow$ & .671$^\uparrow$ & .695$^\uparrow$ &.334$^\uparrow$ & \bf{.595}$^\uparrow$ & .561$^\downarrow$ & \bf{.543}$^\uparrow$ \\
	      & $-$ IEMOCAP  & .619$^\uparrow$ & .634$^\uparrow$ & .641$^\uparrow$ & .696$^\uparrow$ &\bf{.351}$^\uparrow$ & .506$^\downarrow$ & .495$^\downarrow$ & .498$^\uparrow$ \\
	      & $-$ RML    & .544$^\uparrow$ & .631$^\uparrow$ & .648$^\uparrow$ & .611$^\uparrow$ &.309$^\downarrow$ & .560$^\bullet$ & .563$^\downarrow$ & \bf{.547}$^\uparrow$ \\
\hline\\
eNTERFACE     & All           & .455 & .500 & .491 & .468 &\bf{.247} & .484 & \bf{.499} & \bf{.395} \\
	      & $-$ AFEW   & .645$^\uparrow$ & \bf{.694}$^\uparrow$ & \bf{.721}$^\uparrow$ & .737$^\uparrow$ &.238$^\downarrow$ & \bf{.492}$^\uparrow$ & .490$^\downarrow$ & \bf{.397}$^\bullet$ \\
	      & $-$ Emo-DB & .538$^\uparrow$ & .602$^\uparrow$ & .642$^\uparrow$ & .644$^\uparrow$ &.231$^\downarrow$ & \bf{.499}$^\uparrow$ & .482$^\downarrow$ & \bf{.391}$^\bullet$ \\
	      & $-$ EMOVO  & .632$^\uparrow$ & .652$^\uparrow$ & .654$^\uparrow$ & .681$^\uparrow$ &\bf{.248}$^\bullet$ & .476$^\downarrow$ & .477$^\downarrow$ & .375$^\downarrow$ \\
	      & $-$ IEMOCAP  & \bf{.749}$^\uparrow$ & \bf{.696}$^\uparrow$ & \bf{.711}$^\uparrow$ & \bf{.751}$^\uparrow$ &\bf{.244}$^\bullet$ & .481$^\bullet$ & .483$^\downarrow$ & .381$^\downarrow$ \\
	      & $-$ RML    & .624$^\uparrow$ & .625$^\uparrow$ & .639$^\uparrow$ & .674$^\uparrow$ &.233$^\downarrow$ & .471$^\downarrow$ & .472$^\downarrow$ & 384$^\downarrow$ \\
\hline\\
IEMOCAP       & All             & .685 & .685 & .681 & .676 & .441 & \bf{.684} & .672 & \bf{.671} \\
	      & $-$ AFEW      & \bf{.780}$^\uparrow$ & \bf{.762}$^\uparrow$ & \bf{.771}$^\uparrow$ & \bf{.783}$^\uparrow$ & \bf{.470}$^\uparrow$ & \bf{.688}$^\bullet$ & .671$^\bullet$ & .656$^\downarrow$ \\
	      & $-$ Emo-DB    & .739$^\uparrow$ & .722$^\uparrow$ & .737$^\uparrow$ & .741$^\uparrow$ &.435$^\downarrow$ & \bf{.686}$^\bullet$ & .669$^\bullet$ & .649$^\downarrow$ \\
	      & $-$ EMOVO     & .756$^\uparrow$ & .746$^\uparrow$ & .751$^\uparrow$ & .762$^\uparrow$ &.456$^\uparrow$ & \bf{.686}$^\bullet$ & \bf{.680}$^\uparrow$ & \bf{.665}$^\bullet$ \\
	      & $-$ eNTERFACE    & .765$^\uparrow$ & .740$^\uparrow$ & .755$^\uparrow$ & .772$^\uparrow$ &.427$^\downarrow$ & \bf{.680}$^\bullet$ & \bf{.683}$^\uparrow$ & \bf{.667}$^\bullet$ \\
	      & $-$ RML       & .755$^\uparrow$ & .735$^\uparrow$ & .746$^\uparrow$ & .764$^\uparrow$ &.459$^\uparrow$ & .671$^\downarrow$ & \bf{.681}$^\uparrow$ & .659$^\downarrow$ \\
\hline\\
	  RML           & All        & \bf{.615} & .643 & .649 & .626 & .301 & \bf{.669} & \bf{.662} & \bf{.570} \\
	      & $-$ AFEW   & .470$^\downarrow$ & .715$^\uparrow$ & \bf{.744}$^\uparrow$ & \bf{.722}$^\uparrow$ &\bf{.318}$^\uparrow$ & .656$^\downarrow$ & .644$^\downarrow$ & .562$^\downarrow$ \\
	      & $-$ Emo-DB & .503$^\downarrow$ & .709$^\uparrow$ & .714$^\uparrow$ & .663$^\uparrow$ &.297$^\bullet$ & .650$^\downarrow$ & \bf{.664}$^\bullet$ & .543$^\downarrow$ \\
	      & $-$ EMOVO  & .468$^\downarrow$ & .690$^\uparrow$ & .704$^\uparrow$ & .655$^\uparrow$ &.287$^\downarrow$ & .653$^\downarrow$ & .653$^\downarrow$ & .555$^\downarrow$ \\
	      & $-$ eNTERFACE & .471$^\downarrow$ & .710$^\uparrow$ & .707$^\uparrow$ & .687$^\uparrow$ &.302$^\bullet$ & .648$^\downarrow$ & \bf{.660}$^\bullet$ & .557$^\downarrow$ \\
	      & $-$ IEMOCAP  & .543$^\downarrow$ & \bf{.736}$^\uparrow$ & \bf{.754}$^\uparrow$ & .693$^\uparrow$ &.298$^\bullet$ & .656$^\downarrow$ & .656$^\downarrow$ & .533$^\downarrow$ \\
\hline
\end{tabular}
\label{tab:ablation}
\end{table*}

The last set of experiments is concerned with RQ4, that is, to assess the effectiveness of our multi-domain network when compared with state-of-the-art domain adaptation solutions for speech emotion recognition. Table~\ref{tab:sota} shows UAR numbers obtained by Adaptive SVM. The table also shows UAR numbers obtained by our multi-domain network. As can be seen, our multi-domain network outperformed Adaptive SVM in all target domains considered in the study. Gains are statistically significant, and range from 4.3\% to 18.4\%, depending on the target domain.

\begin{table} [htb!]
\caption{UAR numbers for Adaptive SVM and CNN$+$LSTM.}
\centering
\begin{tabular}{lcc|r} \hline
	Target	  & Adaptive SVM & CNN$+$LSTM & Gain\\\hline\\
AFEW      & .539 & .625 & .159 \\
Emo-DB    & .797 & .856 & .074 \\
EMOVO     & .692 & .735 & .062 \\
eNTERFACE & .634 & .751 & .184 \\
IEMOCAP   & .751 & .783 & .043 \\
RML       & .721 & .760 & .054 \\
\hline
  \end{tabular}
\label{tab:sota}
\end{table}

\section{Conclusions}

Automatically recognizing human emotions from speech is currently one of the most challenging tasks in the field of affective computing. In solving this task we are often in the situation that we have a large collection of labeled out-of-domain data but truly desire a model that performs well in a target domain which is short on labeled data. To deal with this situation we proposed a deep architecture which implements a multi-domain network. More specifically,the architecture is a blend of CNN with LSTM networks, and extracts spatial and sequential features from raw audio. In order to evaluate different feature transference approaches, we investigated the best freezing/tuning cut-off for each target domain. We also investigated whether it is beneficial to use target domain data during pre-training. We performed a comprehensive experiment using six domains, which may differ in terms of language, emotions, amount of labels, and recording conditions. Our feature transference approaches provide gains that range from 4.3\% to 18.4\% when compared with recent domain adaptation approaches for speech emotion recognition.

\section{Acknowledgements}

We thank the partial support given by the Brazilian National Institute of Science and Technology for the Web (grant MCT-CNPq 573871/2008-6), Project: Models, Algorithms and Systems for the Web (grant FAPEMIG / PRONEX / MASWeb APQ-01400-14), and authors' individual grants and scholarships from CNPq and Kunumi.

\bibliographystyle{ACM-Reference-Format}
\bibliography{adriano}


\begin{thebibliography}{00}

%%% the Web address.

\ifx \showCODEN    \undefined \def \showCODEN     #1{\unskip}     \fi
\ifx \showDOI      \undefined \def \showDOI       #1{{\tt DOI:}\penalty0{#1}\ }
  \fi
\ifx \showISBNx    \undefined \def \showISBNx     #1{\unskip}     \fi
\ifx \showISBNxiii \undefined \def \showISBNxiii  #1{\unskip}     \fi
\ifx \showISSN     \undefined \def \showISSN      #1{\unskip}     \fi
\ifx \showLCCN     \undefined \def \showLCCN      #1{\unskip}     \fi
\ifx \shownote     \undefined \def \shownote      #1{#1}          \fi
\ifx \showarticletitle \undefined \def \showarticletitle #1{#1}   \fi
\ifx \showURL      \undefined \def \showURL       #1{#1}          \fi
% invisible to TeX
\providecommand\bibfield[2]{#2}
\providecommand\bibinfo[2]{#2}
\providecommand\natexlab[1]{#1}
\providecommand\showeprint[2][]{arXiv:#2}

\bibitem[\protect\citeauthoryear{Abdel{-}Wahab and Busso}{Abdel{-}Wahab and
  Busso}{2015}]%
        {supdomain}
\bibfield{author}{\bibinfo{person}{Mohammed Abdel{-}Wahab} {and}
  \bibinfo{person}{Carlos Busso}.} \bibinfo{year}{2015}\natexlab{}.
\newblock \showarticletitle{Supervised domain adaptation for emotion
  recognition from speech}. In \bibinfo{booktitle}{{\em {IEEE} Intl. Conference
  on Acoustics, Speech and Signal Processing}}. \bibinfo{pages}{5058--5062}.
\newblock


\bibitem[\protect\citeauthoryear{Banziger, Grandjean, and Scherer}{Banziger
  et~al\mbox{.}}{2009}]%
        {psych2}
\bibfield{author}{\bibinfo{person}{Tanja Banziger}, \bibinfo{person}{Didier
  Grandjean}, {and} \bibinfo{person}{Klaus Scherer}.}
  \bibinfo{year}{2009}\natexlab{}.
\newblock \showarticletitle{Emotion recognition from expressions in face,
  voice, and body: the multimodal emotion recognition test (MERT)}.
\newblock \bibinfo{journal}{{\em Emotion\/}} \bibinfo{volume}{9},
  \bibinfo{number}{5} (\bibinfo{year}{2009}), \bibinfo{pages}{691--704}.
\newblock


\bibitem[\protect\citeauthoryear{Batliner, Steidl, Schuller, Seppi, Vogt,
  Wagner, Devillers, Vidrascu, Aharonson, Kessous, and Amir}{Batliner
  et~al\mbox{.}}{2011}]%
        {emotion1}
\bibfield{author}{\bibinfo{person}{Anton Batliner}, \bibinfo{person}{Stefan
  Steidl}, \bibinfo{person}{Bj{\"{o}}rn~W. Schuller}, \bibinfo{person}{Dino
  Seppi}, \bibinfo{person}{Thurid Vogt}, \bibinfo{person}{Johannes Wagner},
  \bibinfo{person}{Laurence Devillers}, \bibinfo{person}{Laurence Vidrascu},
  \bibinfo{person}{Vered Aharonson}, \bibinfo{person}{Lo{\"{\i}}c Kessous},
  {and} \bibinfo{person}{Noam Amir}.} \bibinfo{year}{2011}\natexlab{}.
\newblock \showarticletitle{Whodunnit - Searching for the most important
  feature types signalling emotion-related user states in speech}.
\newblock \bibinfo{journal}{{\em Computer Speech {\&} Language\/}}
  \bibinfo{volume}{25}, \bibinfo{number}{1} (\bibinfo{year}{2011}),
  \bibinfo{pages}{4--28}.
\newblock


\bibitem[\protect\citeauthoryear{Ben{-}David, Blitzer, Crammer, Kulesza,
  Pereira, and Vaughan}{Ben{-}David et~al\mbox{.}}{2010}]%
        {domain5}
\bibfield{author}{\bibinfo{person}{Shai Ben{-}David}, \bibinfo{person}{John
  Blitzer}, \bibinfo{person}{Koby Crammer}, \bibinfo{person}{Alex Kulesza},
  \bibinfo{person}{Fernando Pereira}, {and} \bibinfo{person}{Jennifer~Wortman
  Vaughan}.} \bibinfo{year}{2010}\natexlab{}.
\newblock \showarticletitle{A theory of learning from different domains}.
\newblock \bibinfo{journal}{{\em Machine Learning\/}} \bibinfo{volume}{79},
  \bibinfo{number}{1-2} (\bibinfo{year}{2010}), \bibinfo{pages}{151--175}.
\newblock


\bibitem[\protect\citeauthoryear{Burkhardt, Paeschke, Rolfes, Sendlmeier, and
  Weiss}{Burkhardt et~al\mbox{.}}{2005}]%
        {emodb}
\bibfield{author}{\bibinfo{person}{Felix Burkhardt}, \bibinfo{person}{Astrid
  Paeschke}, \bibinfo{person}{M. Rolfes}, \bibinfo{person}{Walter~F.
  Sendlmeier}, {and} \bibinfo{person}{Benjamin Weiss}.}
  \bibinfo{year}{2005}\natexlab{}.
\newblock \showarticletitle{A database of German emotional speech}. In
  \bibinfo{booktitle}{{\em European Conference on Speech Communication and
  Technology}}. \bibinfo{pages}{1517--1520}.
\newblock


\bibitem[\protect\citeauthoryear{Busso, Bulut, Lee, Kazemzadeh, Mower, Kim,
  Chang, Lee, and Narayanan}{Busso et~al\mbox{.}}{2008}]%
        {iemocap}
\bibfield{author}{\bibinfo{person}{Carlos Busso}, \bibinfo{person}{Murtaza
  Bulut}, \bibinfo{person}{Chi{-}Chun Lee}, \bibinfo{person}{Abe Kazemzadeh},
  \bibinfo{person}{Emily Mower}, \bibinfo{person}{Samuel Kim},
  \bibinfo{person}{Jeannette~N. Chang}, \bibinfo{person}{Sungbok Lee}, {and}
  \bibinfo{person}{Shrikanth Narayanan}.} \bibinfo{year}{2008}\natexlab{}.
\newblock \showarticletitle{{IEMOCAP:} interactive emotional dyadic motion
  capture database}.
\newblock \bibinfo{journal}{{\em Language Resources and Evaluation\/}}
  \bibinfo{volume}{42}, \bibinfo{number}{4} (\bibinfo{year}{2008}),
  \bibinfo{pages}{335--359}.
\newblock


\bibitem[\protect\citeauthoryear{Busso, Parthasarathy, mania, Abdel{-}Wahab,
  Sadoughi, and Provost}{Busso et~al\mbox{.}}{2017}]%
        {corpus}
\bibfield{author}{\bibinfo{person}{Carlos Busso}, \bibinfo{person}{Srinivas
  Parthasarathy}, \bibinfo{person}{Alec~Bur mania}, \bibinfo{person}{Mohammed
  Abdel{-}Wahab}, \bibinfo{person}{Najmeh Sadoughi}, {and}
  \bibinfo{person}{Emily~Mower Provost}.} \bibinfo{year}{2017}\natexlab{}.
\newblock \showarticletitle{{MSP-IMPROV:} an acted corpus of dyadic
  interactions to study emotion perception}.
\newblock \bibinfo{journal}{{\em {IEEE} Transactions on Affective Computing\/}}
  \bibinfo{volume}{8}, \bibinfo{number}{1} (\bibinfo{year}{2017}),
  \bibinfo{pages}{67--80}.
\newblock


\bibitem[\protect\citeauthoryear{Chollet}{Chollet}{2015}]%
        {keras}
\bibfield{author}{\bibinfo{person}{F. Chollet}.}
  \bibinfo{year}{2015}\natexlab{}.
\newblock \bibinfo{title}{keras}.
\newblock \bibinfo{howpublished}{\url{https://github.com/fchollet/keras}}.
  (\bibinfo{year}{2015}).
\newblock


\bibitem[\protect\citeauthoryear{Costantini, Iaderola, Paoloni, and
  Todisco}{Costantini et~al\mbox{.}}{2014}]%
        {emovo}
\bibfield{author}{\bibinfo{person}{Giovanni Costantini},
  \bibinfo{person}{Iacopo Iaderola}, \bibinfo{person}{Andrea Paoloni}, {and}
  \bibinfo{person}{Massimiliano Todisco}.} \bibinfo{year}{2014}\natexlab{}.
\newblock \showarticletitle{{EMOVO} corpus: an Italian emotional speech
  database}. In \bibinfo{booktitle}{{\em Intl. Conference on Language Resources
  and Evaluation}}. \bibinfo{pages}{3501--3504}.
\newblock


\bibitem[\protect\citeauthoryear{Cowie and Cornelius}{Cowie and
  Cornelius}{2003}]%
        {cowie}
\bibfield{author}{\bibinfo{person}{Roddy Cowie} {and}
  \bibinfo{person}{Randolph~R. Cornelius}.} \bibinfo{year}{2003}\natexlab{}.
\newblock \showarticletitle{Describing the emotional states that are expressed
  in speech}.
\newblock \bibinfo{journal}{{\em Speech Communication\/}} \bibinfo{volume}{40},
  \bibinfo{number}{1-2} (\bibinfo{year}{2003}), \bibinfo{pages}{5--32}.
\newblock


\bibitem[\protect\citeauthoryear{Deng, Xu, Zhang, Fr{\"{u}}hholz, and
  Schuller}{Deng et~al\mbox{.}}{2017}]%
        {affect8}
\bibfield{author}{\bibinfo{person}{Jun Deng}, \bibinfo{person}{Xinzhou Xu},
  \bibinfo{person}{Zixing Zhang}, \bibinfo{person}{Sascha Fr{\"{u}}hholz},
  {and} \bibinfo{person}{Bj{\"{o}}rn~W. Schuller}.}
  \bibinfo{year}{2017}\natexlab{}.
\newblock \showarticletitle{Universum autoencoder-based domain adaptation for
  speech emotion recognition}.
\newblock \bibinfo{journal}{{\em {IEEE} Signal Processing Letters\/}}
  \bibinfo{volume}{24}, \bibinfo{number}{4} (\bibinfo{year}{2017}),
  \bibinfo{pages}{500--504}.
\newblock


\bibitem[\protect\citeauthoryear{Deng, Zhang, Eyben, and Schuller}{Deng
  et~al\mbox{.}}{2014b}]%
        {affect4}
\bibfield{author}{\bibinfo{person}{Jun Deng}, \bibinfo{person}{Zixing Zhang},
  \bibinfo{person}{Florian Eyben}, {and} \bibinfo{person}{Bj{\"{o}}rn~W.
  Schuller}.} \bibinfo{year}{2014}\natexlab{b}.
\newblock \showarticletitle{Autoencoder-based unsupervised domain adaptation
  for speech emotion recognition}.
\newblock \bibinfo{journal}{{\em {IEEE} Signal Processing Letters\/}}
  \bibinfo{volume}{21}, \bibinfo{number}{9} (\bibinfo{year}{2014}),
  \bibinfo{pages}{1068--1072}.
\newblock


\bibitem[\protect\citeauthoryear{Deng, Zhang, Marchi, and Schuller}{Deng
  et~al\mbox{.}}{2013}]%
        {baseline}
\bibfield{author}{\bibinfo{person}{Jun Deng}, \bibinfo{person}{Zixing Zhang},
  \bibinfo{person}{Erik Marchi}, {and} \bibinfo{person}{Bj{\"{o}}rn~W.
  Schuller}.} \bibinfo{year}{2013}\natexlab{}.
\newblock \showarticletitle{Sparse Autoencoder-Based Feature Transfer Learning
  for Speech Emotion Recognition}. In \bibinfo{booktitle}{{\em {ACII}
  Association Conference on Affective Computing and Intelligent Interaction}}.
  \bibinfo{pages}{511--516}.
\newblock


\bibitem[\protect\citeauthoryear{Deng, Zhang, and Schuller}{Deng
  et~al\mbox{.}}{2014a}]%
        {domain4}
\bibfield{author}{\bibinfo{person}{Jun Deng}, \bibinfo{person}{Zixing Zhang},
  {and} \bibinfo{person}{Bj{\"{o}}rn~W. Schuller}.}
  \bibinfo{year}{2014}\natexlab{a}.
\newblock \showarticletitle{Linked source and target domain subspace feature
  transfer learning - exemplified by speech emotion recognition}. In
  \bibinfo{booktitle}{{\em Intl. Conference on Pattern Recognition}}.
  \bibinfo{pages}{761--766}.
\newblock


\bibitem[\protect\citeauthoryear{Dhall, Goecke, Lucey, and Gedeon}{Dhall
  et~al\mbox{.}}{2012}]%
        {afew}
\bibfield{author}{\bibinfo{person}{Abhinav Dhall}, \bibinfo{person}{Roland
  Goecke}, \bibinfo{person}{Simon Lucey}, {and} \bibinfo{person}{Tom Gedeon}.}
  \bibinfo{year}{2012}\natexlab{}.
\newblock \showarticletitle{Collecting large, richly annotated
  facial-expression databases from movies}.
\newblock \bibinfo{journal}{{\em {IEEE} MultiMedia\/}} \bibinfo{volume}{19},
  \bibinfo{number}{3} (\bibinfo{year}{2012}), \bibinfo{pages}{34--41}.
\newblock


\bibitem[\protect\citeauthoryear{Drolet, Schubotz, and Fisher}{Drolet
  et~al\mbox{.}}{2012}]%
        {spont}
\bibfield{author}{\bibinfo{person}{Matthis Drolet}, \bibinfo{person}{Ricarda
  Schubotz}, {and} \bibinfo{person}{Julia Fisher}.}
  \bibinfo{year}{2012}\natexlab{}.
\newblock \showarticletitle{Authenticity affects the recognition of emotions in
  speech: Behavioral and fMRI evidence}.
\newblock \bibinfo{journal}{{\em Cognitive Affective \& Behavioral
  Neuroscience\/}} \bibinfo{volume}{12}, \bibinfo{number}{1}
  (\bibinfo{year}{2012}), \bibinfo{pages}{140--150}.
\newblock


\bibitem[\protect\citeauthoryear{Eyben, Schuller, and Rigoll}{Eyben
  et~al\mbox{.}}{2012}]%
        {affect1}
\bibfield{author}{\bibinfo{person}{Florian Eyben},
  \bibinfo{person}{Bj{\"{o}}rn~W. Schuller}, {and} \bibinfo{person}{Gerhard
  Rigoll}.} \bibinfo{year}{2012}\natexlab{}.
\newblock \showarticletitle{Improving generalisation and robustness of acoustic
  affect recognition}. In \bibinfo{booktitle}{{\em Intl. Conference on
  Multimodal Interaction}}. \bibinfo{pages}{517--522}.
\newblock


\bibitem[\protect\citeauthoryear{Han, Yu, and Tashev}{Han
  et~al\mbox{.}}{2014}]%
        {deep1}
\bibfield{author}{\bibinfo{person}{Kun Han}, \bibinfo{person}{Dong Yu}, {and}
  \bibinfo{person}{Ivan Tashev}.} \bibinfo{year}{2014}\natexlab{}.
\newblock \showarticletitle{Speech emotion recognition using deep neural
  network and extreme learning machine}. In \bibinfo{booktitle}{{\em Annual
  Conference of the Intl. Speech Communication Association}}.
  \bibinfo{pages}{223--227}.
\newblock


\bibitem[\protect\citeauthoryear{Huang, Dong, Mao, and Zhan}{Huang
  et~al\mbox{.}}{2014}]%
        {cnn3}
\bibfield{author}{\bibinfo{person}{Zhengwei Huang}, \bibinfo{person}{Ming
  Dong}, \bibinfo{person}{Qirong Mao}, {and} \bibinfo{person}{Yongzhao Zhan}.}
  \bibinfo{year}{2014}\natexlab{}.
\newblock \showarticletitle{Speech emotion recognition using {CNN}}. In
  \bibinfo{booktitle}{{\em {ACM} Intl. Conference on Multimedia}}.
  \bibinfo{pages}{801--804}.
\newblock


\bibitem[\protect\citeauthoryear{Huang, Xue, Mao, and Zhan}{Huang
  et~al\mbox{.}}{2017}]%
        {domain1}
\bibfield{author}{\bibinfo{person}{Zhengwei Huang}, \bibinfo{person}{Wentao
  Xue}, \bibinfo{person}{Qirong Mao}, {and} \bibinfo{person}{Yongzhao Zhan}.}
  \bibinfo{year}{2017}\natexlab{}.
\newblock \showarticletitle{Unsupervised domain adaptation for speech emotion
  recognition using PCANet}.
\newblock \bibinfo{journal}{{\em Multimedia Tools and Applications\/}}
  \bibinfo{volume}{76}, \bibinfo{number}{5} (\bibinfo{year}{2017}),
  \bibinfo{pages}{6785--6799}.
\newblock


\bibitem[\protect\citeauthoryear{Johnstone, van Reekum, Oakes, and
  Davidson}{Johnstone et~al\mbox{.}}{2006}]%
        {neuro4}
\bibfield{author}{\bibinfo{person}{Tom Johnstone}, \bibinfo{person}{Carien van
  Reekum}, \bibinfo{person}{Terrence Oakes}, {and} \bibinfo{person}{Richard
  Davidson}.} \bibinfo{year}{2006}\natexlab{}.
\newblock \showarticletitle{The voice of emotion: an FMRI study of neural
  responses to angry and happy vocal expressions}.
\newblock \bibinfo{journal}{{\em Social Cognitive and Affective
  Neuroscience\/}} \bibinfo{volume}{1}, \bibinfo{number}{3}
  (\bibinfo{year}{2006}), \bibinfo{pages}{242--249}.
\newblock


\bibitem[\protect\citeauthoryear{Juslin and Laukka}{Juslin and Laukka}{2003}]%
        {juslin}
\bibfield{author}{\bibinfo{person}{P. Juslin} {and} \bibinfo{person}{P.
  Laukka}.} \bibinfo{year}{2003}\natexlab{}.
\newblock \showarticletitle{Communication of emotions in vocal expression and
  music performance: Different channels, same code?}
\newblock \bibinfo{journal}{{\em Psychological bulletin\/}}
  \bibinfo{volume}{129}, \bibinfo{number}{5} (\bibinfo{year}{2003}),
  \bibinfo{pages}{770}.
\newblock


\bibitem[\protect\citeauthoryear{Kim, Lee, and Provost}{Kim
  et~al\mbox{.}}{2013}]%
        {deep2}
\bibfield{author}{\bibinfo{person}{Yelin Kim}, \bibinfo{person}{Honglak Lee},
  {and} \bibinfo{person}{Emily~Mower Provost}.}
  \bibinfo{year}{2013}\natexlab{}.
\newblock \showarticletitle{Deep learning for robust feature generation in
  audiovisual emotion recognition}. In \bibinfo{booktitle}{{\em {IEEE} Intl.
  Conference on Acoustics, Speech and Signal Processing}}.
  \bibinfo{pages}{3687--3691}.
\newblock


\bibitem[\protect\citeauthoryear{Koolagudi, Barthwal, Devliyal, and
  Rao}{Koolagudi et~al\mbox{.}}{2012}]%
        {gaussian}
\bibfield{author}{\bibinfo{person}{Shashidhar Koolagudi},
  \bibinfo{person}{Anurag Barthwal}, \bibinfo{person}{Swati Devliyal}, {and}
  \bibinfo{person}{K. Rao}.} \bibinfo{year}{2012}\natexlab{}.
\newblock \showarticletitle{Real life emotion classification from speech using
  gaussian mixture models}. In \bibinfo{booktitle}{{\em Intl. Conference
  Contemporary Computing}}. \bibinfo{pages}{250--261}.
\newblock


\bibitem[\protect\citeauthoryear{Koolagudi and Rao}{Koolagudi and Rao}{2012}]%
        {review}
\bibfield{author}{\bibinfo{person}{Shashidhar Koolagudi} {and}
  \bibinfo{person}{K. Rao}.} \bibinfo{year}{2012}\natexlab{}.
\newblock \showarticletitle{Emotion recognition from speech: a review}.
\newblock \bibinfo{journal}{{\em Intl. Journal of Speech Technology\/}}
  \bibinfo{volume}{15}, \bibinfo{number}{2} (\bibinfo{year}{2012}),
  \bibinfo{pages}{99--117}.
\newblock


\bibitem[\protect\citeauthoryear{Koolagudi, Ray, and Rao}{Koolagudi
  et~al\mbox{.}}{2010}]%
        {rate}
\bibfield{author}{\bibinfo{person}{Shashidhar Koolagudi},
  \bibinfo{person}{Sudhin Ray}, {and} \bibinfo{person}{K. Rao}.}
  \bibinfo{year}{2010}\natexlab{}.
\newblock \showarticletitle{Emotion classification based on speaking rate}. In
  \bibinfo{booktitle}{{\em Intl. Conference Contemporary Computing}}.
  \bibinfo{pages}{316--327}.
\newblock


\bibitem[\protect\citeauthoryear{Mao, Xue, Rao, Zhang, and Zhan}{Mao
  et~al\mbox{.}}{2016}]%
        {domain2}
\bibfield{author}{\bibinfo{person}{Qirong Mao}, \bibinfo{person}{Wentao Xue},
  \bibinfo{person}{Qiyu Rao}, \bibinfo{person}{Feifei Zhang}, {and}
  \bibinfo{person}{Yongzhao Zhan}.} \bibinfo{year}{2016}\natexlab{}.
\newblock \showarticletitle{Domain adaptation for speech emotion recognition by
  sharing priors between related source and target classes}. In
  \bibinfo{booktitle}{{\em {IEEE} Intl. Conference on Acoustics, Speech and
  Signal Processing}}. \bibinfo{pages}{2608--2612}.
\newblock


\bibitem[\protect\citeauthoryear{Marchi, Eyben, Hagerer, and Schuller}{Marchi
  et~al\mbox{.}}{2016}]%
        {affect7}
\bibfield{author}{\bibinfo{person}{Erik Marchi}, \bibinfo{person}{Florian
  Eyben}, \bibinfo{person}{Gerhard Hagerer}, {and}
  \bibinfo{person}{Bj{\"{o}}rn~W. Schuller}.} \bibinfo{year}{2016}\natexlab{}.
\newblock \showarticletitle{Real-time tracking of speakers' emotions, states,
  and traits on mobile platforms}. In \bibinfo{booktitle}{{\em Annual
  Conference of the Intl. Speech Communication Association}}.
  \bibinfo{pages}{1182--1183}.
\newblock


\bibitem[\protect\citeauthoryear{Martin, Kotsia, Macq, and Pitas}{Martin
  et~al\mbox{.}}{2006}]%
        {enterface}
\bibfield{author}{\bibinfo{person}{Olivier Martin}, \bibinfo{person}{Irene
  Kotsia}, \bibinfo{person}{Benoit~M. Macq}, {and} \bibinfo{person}{Ioannis
  Pitas}.} \bibinfo{year}{2006}\natexlab{}.
\newblock \showarticletitle{The eNTERFACE'05 Audio-Visual Emotion Database}. In
  \bibinfo{booktitle}{{\em Intl. Conference on Data Engineering Workshops}}.
  \bibinfo{pages}{8}.
\newblock


\bibitem[\protect\citeauthoryear{Nogueiras, Moreno, Bonafonte, and
  Mari{\~{n}}o}{Nogueiras et~al\mbox{.}}{2001}]%
        {eurospeech}
\bibfield{author}{\bibinfo{person}{Albino Nogueiras},
  \bibinfo{person}{Asunci{\'{o}}n Moreno}, \bibinfo{person}{Antonio Bonafonte},
  {and} \bibinfo{person}{Jos{\'{e}}~B. Mari{\~{n}}o}.}
  \bibinfo{year}{2001}\natexlab{}.
\newblock \showarticletitle{Speech emotion recognition using hidden Markov
  models}. In \bibinfo{booktitle}{{\em European Conference on Speech
  Communication and Technology}}. \bibinfo{pages}{2679--2682}.
\newblock


\bibitem[\protect\citeauthoryear{Oudeyer}{Oudeyer}{2003}]%
        {feat2}
\bibfield{author}{\bibinfo{person}{Pierre{-}Yves Oudeyer}.}
  \bibinfo{year}{2003}\natexlab{}.
\newblock \showarticletitle{The production and recognition of emotions in
  speech: features and algorithms}.
\newblock \bibinfo{journal}{{\em Intl. Journal of Human-Computer Studies\/}}
  \bibinfo{volume}{59}, \bibinfo{number}{1-2} (\bibinfo{year}{2003}),
  \bibinfo{pages}{157--183}.
\newblock


\bibitem[\protect\citeauthoryear{Ramakrishnan and Emary}{Ramakrishnan and
  Emary}{2013}]%
        {review2}
\bibfield{author}{\bibinfo{person}{S. Ramakrishnan} {and}
  \bibinfo{person}{Ibrahiem~El Emary}.} \bibinfo{year}{2013}\natexlab{}.
\newblock \showarticletitle{Speech emotion recognition approaches in human
  computer interaction}.
\newblock \bibinfo{journal}{{\em Telecommunication Systems\/}}
  \bibinfo{volume}{52}, \bibinfo{number}{3} (\bibinfo{year}{2013}),
  \bibinfo{pages}{1467--1478}.
\newblock


\bibitem[\protect\citeauthoryear{Rao, Koolagudi, and Reddy}{Rao
  et~al\mbox{.}}{2013}]%
        {prosodic}
\bibfield{author}{\bibinfo{person}{K. Rao}, \bibinfo{person}{Shashidhar
  Koolagudi}, {and} \bibinfo{person}{Vempada Reddy}.}
  \bibinfo{year}{2013}\natexlab{}.
\newblock \showarticletitle{Emotion recognition from speech using global and
  local prosodic features}.
\newblock \bibinfo{journal}{{\em I. J. Speech Technology\/}}
  \bibinfo{volume}{16}, \bibinfo{number}{2} (\bibinfo{year}{2013}),
  \bibinfo{pages}{143--160}.
\newblock


\bibitem[\protect\citeauthoryear{Sakai}{Sakai}{2014}]%
        {t-test}
\bibfield{author}{\bibinfo{person}{Tetsuya Sakai}.}
  \bibinfo{year}{2014}\natexlab{}.
\newblock \showarticletitle{Statistical reform in information retrieval?}
\newblock \bibinfo{journal}{{\em {SIGIR} Forum\/}} \bibinfo{volume}{48},
  \bibinfo{number}{1} (\bibinfo{year}{2014}), \bibinfo{pages}{3--12}.
\newblock


\bibitem[\protect\citeauthoryear{Schuller, Batliner, Steidl, and
  Seppi}{Schuller et~al\mbox{.}}{2009a}]%
        {emotion4}
\bibfield{author}{\bibinfo{person}{Bj{\"{o}}rn~W. Schuller},
  \bibinfo{person}{Anton Batliner}, \bibinfo{person}{Stefan Steidl}, {and}
  \bibinfo{person}{Dino Seppi}.} \bibinfo{year}{2009}\natexlab{a}.
\newblock \showarticletitle{Emotion recognition from speech: Putting {ASR} in
  the loop}. In \bibinfo{booktitle}{{\em {IEEE} Intl. Conference on Acoustics,
  Speech, and Signal Processing}}. \bibinfo{pages}{4585--4588}.
\newblock


\bibitem[\protect\citeauthoryear{Schuller, Steidl, and Batliner}{Schuller
  et~al\mbox{.}}{2009b}]%
        {inter}
\bibfield{author}{\bibinfo{person}{Bj{\"{o}}rn~W. Schuller},
  \bibinfo{person}{Stefan Steidl}, {and} \bibinfo{person}{Anton Batliner}.}
  \bibinfo{year}{2009}\natexlab{b}.
\newblock \showarticletitle{The {INTERSPEECH} 2009 emotion challenge}. In
  \bibinfo{booktitle}{{\em Annual Conference of the Intl. Speech Communication
  Association}}. \bibinfo{pages}{312--315}.
\newblock


\bibitem[\protect\citeauthoryear{Schuller, Steidl, Batliner, Burkhardt,
  Devillers, M{\"u}ller, Narayanan, et~al\mbox{.}}{Schuller
  et~al\mbox{.}}{2010}]%
        {schuller2010interspeech}
\bibfield{author}{\bibinfo{person}{Bj{\"o}rn~W Schuller},
  \bibinfo{person}{Stefan Steidl}, \bibinfo{person}{Anton Batliner},
  \bibinfo{person}{Felix Burkhardt}, \bibinfo{person}{Laurence Devillers},
  \bibinfo{person}{Christian~A M{\"u}ller}, \bibinfo{person}{Shrikanth~S
  Narayanan}, {and} \bibinfo{person}{others}.} \bibinfo{year}{2010}\natexlab{}.
\newblock \showarticletitle{The INTERSPEECH 2010 paralinguistic challenge.}. In
  \bibinfo{booktitle}{{\em Interspeech}}, Vol.~\bibinfo{volume}{2010}.
  \bibinfo{pages}{2795--2798}.
\newblock


\bibitem[\protect\citeauthoryear{Schuller, Vlasenko, Eyben, W{\"{o}}llmer,
  Stuhlsatz, Wendemuth, and Rigoll}{Schuller et~al\mbox{.}}{2015}]%
        {affect6}
\bibfield{author}{\bibinfo{person}{Bj{\"{o}}rn~W. Schuller},
  \bibinfo{person}{Bogdan Vlasenko}, \bibinfo{person}{Florian Eyben},
  \bibinfo{person}{Martin W{\"{o}}llmer}, \bibinfo{person}{Andr{\'{e}}
  Stuhlsatz}, \bibinfo{person}{Andreas Wendemuth}, {and}
  \bibinfo{person}{Gerhard Rigoll}.} \bibinfo{year}{2015}\natexlab{}.
\newblock \showarticletitle{Cross-corpus acoustic emotion recognition:
  Variances and strategies}. In \bibinfo{booktitle}{{\em Intl. Conference on
  Affective Computing and Intelligent Interaction}}. \bibinfo{pages}{470--476}.
\newblock


\bibitem[\protect\citeauthoryear{Seppi, Batliner, Schuller, Steidl, Vogt,
  Wagner, Devillers, Vidrascu, Amir, and Aharonson}{Seppi
  et~al\mbox{.}}{2008}]%
        {emotion3}
\bibfield{author}{\bibinfo{person}{Dino Seppi}, \bibinfo{person}{Anton
  Batliner}, \bibinfo{person}{Bj{\"{o}}rn~W. Schuller}, \bibinfo{person}{Stefan
  Steidl}, \bibinfo{person}{Thurid Vogt}, \bibinfo{person}{Johannes Wagner},
  \bibinfo{person}{Laurence Devillers}, \bibinfo{person}{Laurence Vidrascu},
  \bibinfo{person}{Noam Amir}, {and} \bibinfo{person}{Vered Aharonson}.}
  \bibinfo{year}{2008}\natexlab{}.
\newblock \showarticletitle{Patterns, prototypes, performance: classifying
  emotional user states}. In \bibinfo{booktitle}{{\em Annual Conference of the
  Intl. Speech Communication Association}}. \bibinfo{pages}{601--604}.
\newblock


\bibitem[\protect\citeauthoryear{Song, Zheng, Ou, Zhang, Jin, Liu, and Yu}{Song
  et~al\mbox{.}}{2016}]%
        {cross3}
\bibfield{author}{\bibinfo{person}{Peng Song}, \bibinfo{person}{Wenming Zheng},
  \bibinfo{person}{Shifeng Ou}, \bibinfo{person}{Xinran Zhang},
  \bibinfo{person}{Yun Jin}, \bibinfo{person}{Jinglei Liu}, {and}
  \bibinfo{person}{Yanwei Yu}.} \bibinfo{year}{2016}\natexlab{}.
\newblock \showarticletitle{Cross-corpus speech emotion recognition based on
  transfer non-negative matrix factorization}.
\newblock \bibinfo{journal}{{\em Speech Communication\/}}  \bibinfo{volume}{83}
  (\bibinfo{year}{2016}), \bibinfo{pages}{34--41}.
\newblock


\bibitem[\protect\citeauthoryear{Spreckelmeyer, Kutas, Urbach, Altenmuller, and
  Munte}{Spreckelmeyer et~al\mbox{.}}{2009}]%
        {neuro3}
\bibfield{author}{\bibinfo{person}{Katja Spreckelmeyer}, \bibinfo{person}{Marta
  Kutas}, \bibinfo{person}{Thomas Urbach}, \bibinfo{person}{Eckart
  Altenmuller}, {and} \bibinfo{person}{Thomas Munte}.}
  \bibinfo{year}{2009}\natexlab{}.
\newblock \showarticletitle{Neural processing of vocal emotion and identity}.
\newblock \bibinfo{journal}{{\em Brain and Cognition\/}} \bibinfo{volume}{69},
  \bibinfo{number}{1} (\bibinfo{year}{2009}), \bibinfo{pages}{121--126}.
\newblock


\bibitem[\protect\citeauthoryear{Stienen, Tanaka, and de~Gelder}{Stienen
  et~al\mbox{.}}{2011}]%
        {neuro2}
\bibfield{author}{\bibinfo{person}{B. Stienen}, \bibinfo{person}{A. Tanaka},
  {and} \bibinfo{person}{B. de Gelder}.} \bibinfo{year}{2011}\natexlab{}.
\newblock \showarticletitle{Emotional voice and emotional body postures
  influence each other independently of visual awareness}.
\newblock \bibinfo{journal}{{\em Plos One\/}} \bibinfo{volume}{10},
  \bibinfo{number}{6} (\bibinfo{year}{2011}), \bibinfo{pages}{e25517}.
\newblock


\bibitem[\protect\citeauthoryear{Stuhlsatz, Meyer, Eyben, Zielke, Meier, and
  Schuller}{Stuhlsatz et~al\mbox{.}}{2011}]%
        {affect2}
\bibfield{author}{\bibinfo{person}{Andr{\'{e}} Stuhlsatz},
  \bibinfo{person}{Christine Meyer}, \bibinfo{person}{Florian Eyben},
  \bibinfo{person}{Thomas Zielke}, \bibinfo{person}{Hans{-}G{\"{u}}nter Meier},
  {and} \bibinfo{person}{Bj{\"{o}}rn~W. Schuller}.}
  \bibinfo{year}{2011}\natexlab{}.
\newblock \showarticletitle{Deep neural networks for acoustic emotion
  recognition: Raising the benchmarks}. In \bibinfo{booktitle}{{\em {IEEE}
  Intl. Conference on Acoustics, Speech, and Signal Processing}}.
  \bibinfo{pages}{5688--5691}.
\newblock


\bibitem[\protect\citeauthoryear{Tanaka, Koizumi, Imai, Hiramatsu, Hiramoto,
  and de~Gelder}{Tanaka et~al\mbox{.}}{2010}]%
        {neuro1}
\bibfield{author}{\bibinfo{person}{A. Tanaka}, \bibinfo{person}{A. Koizumi},
  \bibinfo{person}{H. Imai}, \bibinfo{person}{S. Hiramatsu},
  \bibinfo{person}{E. Hiramoto}, {and} \bibinfo{person}{B. de Gelder}.}
  \bibinfo{year}{2010}\natexlab{}.
\newblock \showarticletitle{I feel your voice: Cultural differences in the
  multisensory perception of emotion}.
\newblock \bibinfo{journal}{{\em Psychological Science\/}}
  \bibinfo{volume}{21}, \bibinfo{number}{9} (\bibinfo{year}{2010}),
  \bibinfo{pages}{1259--1262}.
\newblock


\bibitem[\protect\citeauthoryear{Vincent, Larochelle, Bengio, and
  Manzagol}{Vincent et~al\mbox{.}}{2008}]%
        {autoencoder}
\bibfield{author}{\bibinfo{person}{Pascal Vincent}, \bibinfo{person}{Hugo
  Larochelle}, \bibinfo{person}{Yoshua Bengio}, {and}
  \bibinfo{person}{Pierre{-}Antoine Manzagol}.}
  \bibinfo{year}{2008}\natexlab{}.
\newblock \showarticletitle{Extracting and composing robust features with
  denoising autoencoders}. In \bibinfo{booktitle}{{\em Intl. Conference on
  Machine Learning}}. \bibinfo{pages}{1096--1103}.
\newblock


\bibitem[\protect\citeauthoryear{Williams and Stevens}{Williams and
  Stevens}{1972}]%
        {feat3}
\bibfield{author}{\bibinfo{person}{Carl Williams} {and}
  \bibinfo{person}{Kenneth Stevens}.} \bibinfo{year}{1972}\natexlab{}.
\newblock \showarticletitle{Emotions and speech: some acoustical correlates}.
\newblock \bibinfo{journal}{{\em The Journal of the Acoustical Society of
  America\/}} \bibinfo{volume}{52}, \bibinfo{number}{4} (\bibinfo{year}{1972}),
  \bibinfo{pages}{1238--1250}.
\newblock


\bibitem[\protect\citeauthoryear{W{\"{o}}llmer, Kaiser, Eyben, Schuller, and
  Rigoll}{W{\"{o}}llmer et~al\mbox{.}}{2013}]%
        {affect3}
\bibfield{author}{\bibinfo{person}{Martin W{\"{o}}llmer},
  \bibinfo{person}{Moritz Kaiser}, \bibinfo{person}{Florian Eyben},
  \bibinfo{person}{Bj{\"{o}}rn~W. Schuller}, {and} \bibinfo{person}{Gerhard
  Rigoll}.} \bibinfo{year}{2013}\natexlab{}.
\newblock \showarticletitle{LSTM-Modeling of continuous emotions in an
  audiovisual affect recognition framework}.
\newblock \bibinfo{journal}{{\em Image and Vision Computing\/}}
  \bibinfo{volume}{31}, \bibinfo{number}{2} (\bibinfo{year}{2013}),
  \bibinfo{pages}{153--163}.
\newblock


\bibitem[\protect\citeauthoryear{Xue, Huang, Luo, and Mao}{Xue
  et~al\mbox{.}}{2015}]%
        {dd}
\bibfield{author}{\bibinfo{person}{Went Xue}, \bibinfo{person}{Zhengwei Huang},
  \bibinfo{person}{Xin Luo}, {and} \bibinfo{person}{Qirong Mao}.}
  \bibinfo{year}{2015}\natexlab{}.
\newblock \showarticletitle{Learning speech emotion features by joint
  disentangling-discrimination}. In \bibinfo{booktitle}{{\em Intl. Conference
  on Affective Computing and Intelligent Interaction}}.
  \bibinfo{pages}{374--379}.
\newblock


\bibitem[\protect\citeauthoryear{Yosinski, Clune, Bengio, and Lipson}{Yosinski
  et~al\mbox{.}}{2014}]%
        {bengio2}
\bibfield{author}{\bibinfo{person}{Jason Yosinski}, \bibinfo{person}{Jeff
  Clune}, \bibinfo{person}{Yoshua Bengio}, {and} \bibinfo{person}{Hod Lipson}.}
  \bibinfo{year}{2014}\natexlab{}.
\newblock \showarticletitle{How transferable are features in deep neural
  networks?}. In \bibinfo{booktitle}{{\em Annual Conference on Neural
  Information Processing Systems}}. \bibinfo{pages}{3320--3328}.
\newblock


\bibitem[\protect\citeauthoryear{Zhang, Provost, and Essl}{Zhang
  et~al\mbox{.}}{2016a}]%
        {cross2}
\bibfield{author}{\bibinfo{person}{Biqiao Zhang}, \bibinfo{person}{Emily~Mower
  Provost}, {and} \bibinfo{person}{Georg Essl}.}
  \bibinfo{year}{2016}\natexlab{a}.
\newblock \showarticletitle{Cross-corpus acoustic emotion recognition from
  singing and speaking: {A} multi-task learning approach}. In
  \bibinfo{booktitle}{{\em {IEEE} Intl. Conference on Acoustics, Speech and
  Signal Processing}}. \bibinfo{pages}{5805--5809}.
\newblock


\bibitem[\protect\citeauthoryear{Zhang, Ringeval, Han, Deng, Marchi, and
  Schuller}{Zhang et~al\mbox{.}}{2016b}]%
        {autoencoder3}
\bibfield{author}{\bibinfo{person}{Zixing Zhang}, \bibinfo{person}{Fabien
  Ringeval}, \bibinfo{person}{Jing Han}, \bibinfo{person}{Jun Deng},
  \bibinfo{person}{Erik Marchi}, {and} \bibinfo{person}{Bj{\"{o}}rn~W.
  Schuller}.} \bibinfo{year}{2016}\natexlab{b}.
\newblock \showarticletitle{Facing realism in spontaneous emotion recognition
  from speech: feature enhancement by autoencoder with {LSTM} neural networks}.
  In \bibinfo{booktitle}{{\em Annual Conference of the Intl. Speech
  Communication Association}}. \bibinfo{pages}{3593--3597}.
\newblock


\end{thebibliography}

\end{document}